\documentclass[nopacs,epj]{svjour}
\usepackage{graphicx}

\usepackage{amssymb}
\usepackage{amsmath}

\usepackage{multirow}

\usepackage[ps2pdf,colorlinks]{hyperref}

\AtBeginDocument{%
    \hypersetup{
      urlcolor     = blue, 
      linkcolor    = black, 
      citecolor   = blue, 
      pdfborder={0 0 0},
      urlbordercolor={1 1 1},
      citebordercolor={1 1 1}
    }
    }

\begin{document}

\hyphenation{coun-ter cor-res-pon-din-gly e-xam-ple
co-o-pe-ra-tion ex-pe-ri-men-tal}

\title{Complementarity, quantum erasure and delayed choice with modified Mach-Zehnder interferometers}
\author{Stefan~Ataman
}                     
%
%
\institute{ECE Paris, \email{ataman@ece.fr}}
\date{Received: date / Revised version: date}
%
\abstract{Often cited dictums in Quantum Mechanics include
``observation disturbance causes loss of interference'' and
``ignorance is interference''. In this paper we propose and
describe a series of experiments with modified Mach-Zehnder
interferometers showing that one has to be careful when applying
such dictums. We are able to show that without interacting in any
way with the light quantum (or quanta) expected to behave
``wave-like'', interference fringes can be lost by simply gaining
(or having the potential to gain) the which-path knowledge.
Erasing this information may revive the interference fringes.
Delayed choice can be added, arriving to an experiment in line
with Wheeler's original proposal. We also show that ignorance is
not always synonym with having the interference fringes. The
often-invoked ``collapse of the wavefunction'' is found to be a
non-necessary ingredient to describe our experiments.
%
\PACS{
      {PACS-key}{discribing text of that key}   \and
      {PACS-key}{discribing text of that key}
     } 
} 
%


\maketitle
%

\section{Introduction}

The famous Einstein-Bohr debates \cite{Boh49} and Einstein's
counter-examples to the then-newly founded Quantum Mechanics treat
the loss of interference in a two-slit experiment as a consequence
of the perturbation induced by the measuring device. Heisenberg
used this fact in his famous \emph{Gedankenexperiment}
\cite{Hei27} when he introduced the uncertainty principle: a
disturbance based analysis for the observation of an electron in
orbit with a (gamma ray) light quantum. Bohr \cite{Boh28} was the
first to realize that Quantum Mechanics is more subtle than this
``observation caused disturbance induces the uncertainty'' dictum
and his Complementary Principle refined the so-called
wave-particle duality to a new level. This work spurred various
theoretical refinements \cite{Eng96,Eng95} and experimental
verifications \cite{Cha95,Eic93,Dur98,Ber01,Auc12}.

A couple of decades after Bohr's proposal, Wheeler
\cite{Whe78,Whe83} introduced the idea of delayed-choice
experiment. One could decide what type of phenomenon is measured
\emph{after} the actual measurement was done. In his original
\emph{Gedankenexperiment} one could decide to remove or not the
second beam splitter in a Mach-Zehnder interferometer after the
first light quantum left the first beam splitter, this deciding on
the wave-like or particle-like phenomenon that is measured. It
took several decades until Jacques \emph{et al.}
\cite{Jac07,Jac08} fully tested Wheeler's original delayed choice
proposal. Previous experiments tested equivalent schemes
\cite{Hel87,Bal89}.

Meanwhile the new idea of \emph{Quantum Eraser} was proposed by
Scully and Dr\"uhl \cite{Scu82}. In fact, it was possible to
``erase'' a which path information and -- surprisingly -- revive
the interference fringes previously washed out by the which-path
markers \cite{Scu91}. This proposition stirred also an interesting
controversy \cite{Moh96,Eng99,Moh99}, showing how difficult to
grasp the predictions of Quantum Mechanics can be. A detailed
discussion of the initial proposition can be found in reference
\cite{Scu98}.


The spectacular evolutions in Quantum Optics (QO) made available
sources of single and twin light quanta (via the process of
Spontaneous Parametric Down-Conversions - SPDC
\cite{Har67,Bur70}). Therefore, many fundamental experiments from
Quantum Mechanics became feasible in this field (for a review, see
for example Steinberg \emph{et al.} \cite{Ste96}). The so-called
``high-NOON'' states \emph{i.e.} quantum states of the type
$\vert\psi\rangle=1/\sqrt{2}\left(\vert{N}0\rangle+\vert0N\rangle\right)$
became an experimental reality \cite{Mit04,Kim09,Afe10,Su14} in
the last decade. We shall consider this type of maximally
entangled quantum states in our experimental setup.

The Mach-Zehnder interferometer (MZI) is often chosen to test
various crucial features of the quantum nature of light. Its
versatility led to its use in many experiments
\cite{Gra86,Rar90,Fra91,Ou90,EV93,Kwi95,Shi94}. With a single
light quantum at one input (the other one being kept ``dark''),
the rate of photo-detection at any of its outputs oscillates as
the path-length difference of the interferometer is swept
\cite{Gra86} (and obviously no coincident counts are detected).
Applying pairs of light quanta at its inputs (\emph{i.e.} allowing
fourth-order correlation measurements), shows quantum phenomena
impossible to explain with a semi-classical theory. Having two
light quanta in the MZI implies also that the spatial frequency of
the output interference fringes doubles \cite{Rar90,Shi94}.

Ou \emph{et al.} \cite{Ou90b} inserted parametric down-converters
in the arms of a MZI and showed that the coincidence counts
between the inner and the outer MZI show a sinusoidal variation as
the pump phases are varied. We shall make reference to their work
later on. Walborn \emph{et al.} \cite{Wal02} and later Kwiat
\emph{et al.} \cite{Kwi92} proposed fully-fledged optical quantum
erasers, where the interference can be destroyed and revived in
function of the measured polarization of the coincident choices.
Y-H. Kim \emph{et al.} \cite{Kim00} used a modified Mach-Zehnder
interferometer to create an optical version of the delayed-choice
quantum eraser. By choosing among the possible coincident
detections, one can select the which-path information and lose the
interference or, erase this information and regain the
interference fringes. A theoretical proposal for a quantum eraser
experiment using two MZIs connected via Kerr crystals was done by
Hong \emph{et al.} \cite{Hon08}.

In this paper, we propose and describe in the formalism of QO
several experiments able to show both delayed choice and quantum
erasure. Contrary to other experiments, we use a maximally
entangled state inside the Mach-Zehnder interferometer followed by
the detection of one (of two or more) photons as which-path
marker. We show that as soon as we got the which-way information
(in German \emph{welcher Weg}), the interference fringes disappear
although we \emph{did not interact} in any way with the light
quantum (or quanta) supposed to interfere in the MZI. Erasing this
information brings back the interference fringes, and since this
decision can be done with a space-like separation from the other
detection event, we obtain a delayed-choice experiment in line
with Wheeler's original proposal.

The ``collapse of the wavefunction'' picture is also questioned.
Bohr's complementarity principle is shown to fully explain all
expected experimental results and the picture of ``collapse'' is
found to be merely a mental process of a \emph{particular
experimenter} gaining \emph{some particular information} about the
measured quantum system.

This paper is organized as follows. In Section
\ref{sec:braced_MZIs_particle_nature} we describe the first
experiment able to show the particle nature of light. In Section
\ref{sec:braced_MZIs_particle_wave}, the which-path information is
erased and therefore we are able to show the wave nature of light
quanta.  All delayed choice scenarios and their implications are
discussed in detail in Section
\ref{sec:delayed_choice_quantum_eraser}. The idea of
``engineered'' input states showing no interference fringes in
both previously discussed experimental setups is introduced in
Section \ref{sec:engineered_states}. An extension to our
experimental setup to three or more braced MZIs is discussed in
Section \ref{sec:Wigner_s_friend}. Finally, conclusions are drawn
in Section \ref{sec:conclusions}.


\section{A Mach-Zehnder interferometer with which-path knowledge}
\label{sec:braced_MZIs_particle_nature} 
\begin{figure}
\centering
\includegraphics[width=2in]{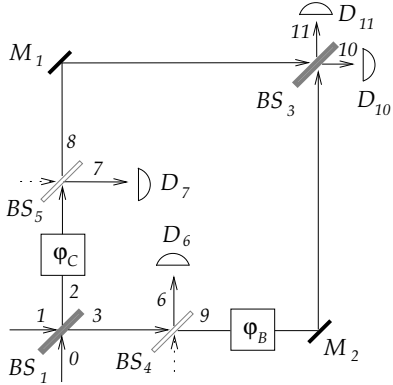}
\caption{To the Mach-Zehnder interferometer formed by
 the beam splitters $\text{BS}_1$ and $\text{BS}_3$, together with the mirrors
$\text{M}_1$ and $\text{M}_2$ a beam splitter was introduced in
each arm of the interferometer (denoted $\text{BS}_4$ and,
respectively, $\text{BS}_5$). We can have thus the which-path
information if any of $D_6$ or $D_7$ ``clicks''. The delays
$\varphi_C$ and $\varphi_B$ allow one to modify the path length
difference in the MZI.}
\label{fig:Braced_MZIs_particle_nature_experiment}
\end{figure}

We consider the
experimental setup depicted in
Fig.~\ref{fig:Braced_MZIs_particle_nature_experiment}. The
Mach-Zehnder interferometer is composed of the beam splitters
$\text{BS}_1$ and $\text{BS}_3$, together with the mirrors
$\text{M}_1$ and $\text{M}_2$. The beam splitters are assumed
identical and are characterized by the transmission (reflection),
coefficients $T$ ($R$). However, in each arm of the interferometer
we introduced a beam splitter (denoted $\text{BS}_4$ and,
respectively, $\text{BS}_5$). These two newly introduced beam
splitters (assumed identical) are characterized by the
coefficients $T_1$ (transmission) and $R_1$ (reflection). The
delays $\varphi_C$ and $\varphi_B$ are voluntarily introduced in
the upper and, respectively, lower path of the interferometer. We
further assume that with the two delays set to zero the length of
the two paths of the MZI are equal. Detectors $D_{10}$ and
$D_{11}$ are placed at the two outputs of beam splitter
$\text{BS}_3$. Throughout this paper, we shall assume ideal
photo-detectors. The notation used to describe this experiment was
done for future convenience (see Section
\ref{sec:braced_MZIs_particle_wave}).

At the inputs labelled ``0'' and ``1'' we apply pairs of identical
photons, \emph{i.e.} the input state can be written as
\begin{equation}
\label{eq:psi_input_state_11}
\vert\psi_{in}\rangle=\vert1_01_1\rangle=\hat{a}_0^\dagger\hat{a}_1^\dagger\vert0\rangle
\end{equation}
where $\hat{a}_k^\dagger$ denotes the creation operator for the
mode (port) $k$. The state $\vert1_01_1\rangle$ denotes a Fock
state with one light quantum in both ports $0$ and $1$ and
$\vert0\rangle$ denotes the vacuum state. The input field
operators obey the usual commutation relations
$[\hat{a}_l,\hat{a}_k]=[\hat{a}_l^\dagger,\hat{a}_k^\dagger]=0$
and $[\hat{a}_l,\hat{a}_k^\dagger]=\delta_{lk}$ where
$\delta_{lk}$ is the Kronecker delta and $l,k=0,1$. Imposing the
same commutation relations to the output field operators, one ends
up with the well-known constraints \cite{Lou03,Ger04,Fea87} on the
beam splitter
\begin{equation}
\vert{T}\vert^2+\vert{R}\vert^2=1
\end{equation}
and
\begin{equation}
RT^*+TR^*=0
\end{equation}
Using the transformation equations for the creation operators
\begin{equation}
\label{eq:a0_dagger_in_respect_w_a2_a3_BS}
\hat{a}_0^\dagger=T\hat{a}_2^\dagger+R\hat{a}_3^\dagger
\end{equation}
and
\begin{equation}
\label{eq:a1_dagger_in_respect_w_a2_a3_BS}
\hat{a}_1^\dagger=R\hat{a}_2^\dagger+T\hat{a}_3^\dagger
\end{equation}
we easily find the state vector after $\text{BS}_1$, namely
\begin{equation}
\label{eq:BS_antibunching} \vert\psi_{23}\rangle
=\sqrt{2}TR\left(\vert0_22_3\rangle+\vert2_20_3\rangle\right)+(T^2+R^2)\vert1_21_3\rangle
\end{equation}
Throughout this paper, when dealing with a balanced (50/50) beam
splitter, we shall use $T=1/\sqrt{2}$ and $R=i/\sqrt{2}$
\cite{Lou03} implying $R^2+T^2=0$, therefore in
Eq.~\eqref{eq:BS_antibunching} the $\vert1_21_3\rangle$ output
state vanishes yielding
\begin{equation}
\label{eq:state_vector_after_BS1_balanced}
\vert\psi_{23}\rangle=\frac{i}{\sqrt{2}}\left(\vert0_22_3\rangle+\vert2_20_3\rangle\right)
\end{equation}
This is called the \emph{antibunching} or HOM effect
\cite{Lou03,HOM87,Fea89}. For the experiment under discussion,
Eq.~\eqref{eq:state_vector_after_BS1_balanced} describes a crucial
point: in our interferometer, the light quanta will ``bunch''
\emph{i.e.} if one light quantum is detected by, say, detector
$\text{D}_6$, we know \emph{with certainty} that the other light
quantum is in the lower arm of the interferometer. Thus, we have
the \emph{welcher Weg} information. However, we got this
information without interacting with (or disturbing) \emph{in any
way} the second light quantum.

In the following, we will describe the proposed experiment using
the standard formalism of QO. Using the fact that an input vacuum
state transforms into an output vacuum state, one can compute the
output state vector. We first write the input field operators
$\hat{a}_0^\dagger$ and $\hat{a}_1^\dagger$ in respect with the
output ones. We obtain
\begin{eqnarray}
\label{eq:a0_dagger_in_respect_w_a6_a7_a10_a11_particle}
\hat{a}_0^\dagger=T_1\left(T^2\text{e}^{i\varphi_C}+R^2\text{e}^{i\varphi_B}\right)\hat{a}_{10}^\dagger
\qquad\qquad\qquad\qquad\quad
\nonumber\\ 
+T_1TR\left(\text{e}^{i\varphi_C}+\text{e}^{i\varphi_B}\right)\hat{a}_{11}^\dagger
+R_1R\hat{a}_{6}^\dagger+R_1T\text{e}^{i\varphi_C}\hat{a}_{7}^\dagger
\end{eqnarray}
and
\begin{eqnarray}
\label{eq:a1_dagger_in_respect_w_a6_a7_a10_a11_particle}
\hat{a}_1^\dagger=T_1TR\left(\text{e}^{i\varphi_C}+\text{e}^{i\varphi_B}\right)\hat{a}_{10}^\dagger
\qquad\qquad\qquad\qquad\qquad
\nonumber\\ 
+T_1\left(T^2\text{e}^{i\varphi_B}+R^2\text{e}^{i\varphi_C}\right)\hat{a}_{11}^\dagger
+R_1T\hat{a}_{6}^\dagger+R_1R\text{e}^{i\varphi_C}\hat{a}_{7}^\dagger
\end{eqnarray}
Assuming the beam splitters $\text{BS}_1$ and $\text{BS}_3$ to be
balanced (50/50), after some calculations, one gets the
output\footnote{Strictly speaking we should have written in
Eq.~\eqref{eq:psi_output_state_particle_nature}
$\vert0_{6}0_{7}2_{10}0_{11}\rangle$ instead of
$\vert2_{10}0_{11}\rangle$, $\vert0_{6}0_{7}0_{10}2_{11}\rangle$
instead of $\vert0_{10}2_{11}\rangle$ etc. In order to keep the
notation simple, we preferred to denote explicitly only the two
modes that appeared in the calculation of the respective terms.}
state
\begin{eqnarray}
\label{eq:psi_output_state_particle_nature}
\vert\psi_{out}\rangle=
\frac{T_1^2\text{e}^{i(\varphi_C+\varphi_B)}}{\sqrt{2}}\sin\left(\Delta\varphi_B\right)
\Big(\vert2_{10}0_{11}\rangle-\vert0_{10}2_{11}\rangle\Big)
\nonumber\\
-T_1^2\text{e}^{i(\varphi_C+\varphi_B)}\cos\left(\Delta\varphi_B\right)\vert1_{10}1_{11}\rangle
\nonumber\\ 
+\frac{iR_1^2}{\sqrt{2}}\Big(\vert2_{6}0_{7}\rangle+\text{e}^{i\varphi_C}\vert0_{6}2_{7}\rangle\Big)
+\frac{T_1R_1}{\sqrt{2}}\Big(-\text{e}^{i\varphi_B}\vert1_{6}1_{10}\rangle
\nonumber\\ 
+i\text{e}^{i\varphi_C}\vert1_{7}1_{10}\rangle
+i\text{e}^{i\varphi_B}\vert1_{6}1_{11}\rangle-\text{e}^{i\varphi_C}\vert1_{7}1_{11}\rangle\Big)
\quad
\end{eqnarray}
where we denoted $\Delta\varphi_B=\varphi_B-\varphi_C$. Using
Eq.~\eqref{eq:psi_output_state_particle_nature} allows us to
compute various detection probabilities at the output ports. For
example, the probability of coincident counts at detectors
$D_{10}$ and $D_{11}$ is given by
\begin{eqnarray}
\label{eq:P_coinc_10_11_particle}
P_{10-11}=\vert\langle1_{10}1_{11}\vert\psi_{out}\rangle\vert^2=
\vert{T}_1\vert^4\cos^2\left(\Delta\varphi_B\right)
\end{eqnarray}
showing indeed the interference pattern frequency expected from a
MZI with two simultaneously impinging light quanta at its input
\cite{Rar90}. Since none of the detectors $D_6$ and $D_7$
``clicked'', we have no which-path information, therefore this
interference pattern should come as no surprise\footnote{It is
noteworthy that the coincidence probability at detectors $D_6$ and
$D_7$ is zero. At a closer look, this is simply the HOM or
antibunching effect.}.

One could try to spy on the \emph{welcher Weg} information by
monitoring the coincident counts between one of the inner
detectors ($D_6$ or $D_7$) and one of the outer ones ($D_{10}$ or
$D_{11}$). If we consider a detection event at detector $D_6$, we
could naively assume that after $\text{BS}_4$ and $\text{BS}_5$
the quantum state inside the MZI is
$1/\sqrt{2}(\vert1_80_9\rangle+\text{e}^{i\varphi_C}\vert0_81_9\rangle)$,
therefore we expect interference fringes on checking the counts at
say, $D_{10}$ (conditioned on a detection at $D_{6}$) similar to
other experiments using a MZI with a single quantum of light at
its input \cite{Gra86}. But this is not what we find if we project
the output state Eq.~\eqref{eq:psi_output_state_particle_nature}
onto the state $\vert1_{6}1_{10}\rangle$. Indeed, we get
\begin{eqnarray}
\label{eq:P_coinc_6_10_particle}
P_{6-10}=\vert\langle1_{6}1_{10}\vert\psi_{out}\rangle\vert^2=
\frac{\vert{T}_1R_1\vert^2}{2}
\end{eqnarray}
and we find no interference fringes on varying the
interferometer's arm length difference. At a more careful look,
this should come at no surprise since by detecting the first light
quantum at the detector $D_6$ we got the \emph{which-path
information} for the second light quantum: it took, \emph{with
certainty}, the lower arm of the interferometer. Therefore, we
could assume that the state inside the MZI after this detection
``collapsed'' to $\vert0_81_9\rangle$. Or, we can simply take the
piece of information ``detection event at $D_6$'' and
correspondingly perform a state reduction yielding the same
result.

As discussed in Appendix \ref{app:density_matrix}, by simply
ignoring the inner detectors $D_6$ and $D_7$ and focusing on the
outer ones, we get no additional information.

One might wonder if the interference fringes could be somehow
restored while still reading a detection at $D_6$ (or $D_7$). The
answer is affirmative if we modify the experimental setup, so that
the which-path information given by these detectors is
\emph{erased}. For example, if in
Eq.~\eqref{eq:P_coinc_6_10_particle} instead of projecting the
state vector $\vert\psi_{out}\rangle$ onto
$\vert1_{6}1_{10}\rangle$ we compute its projection onto
$1/\sqrt{2}\left(\vert1_{6}1_{10}\rangle-i\vert1_{7}1_{10}\rangle\right)$,
we would get
\begin{eqnarray}
\label{eq:P_coinc_6_10_particle_erased} P_{6-10}'
=\Big\vert\frac{1}{\sqrt{2}}\left(\langle1_{6}1_{10}\vert+i\langle1_{7}1_{10}\vert\right)\psi_{out}\rangle\Big\vert^2
\nonumber\\ 
 \sim \cos^2\left(\frac{\Delta\varphi_B}{2}\right)
\end{eqnarray}
and the interference fringes are revived. This is equivalent to
inserting a new beam splitter between detectors $D_6$ and $D_7$
and in the paths coming from the beam splitters $\text{BS}_4$ and
$\text{BS}_5$. This remark takes us to the next experimental
setup.

\section{Two braced Mach-Zehnder interferometers}
\label{sec:braced_MZIs_particle_wave} 
\begin{figure}
\centering
\includegraphics[width=2in]{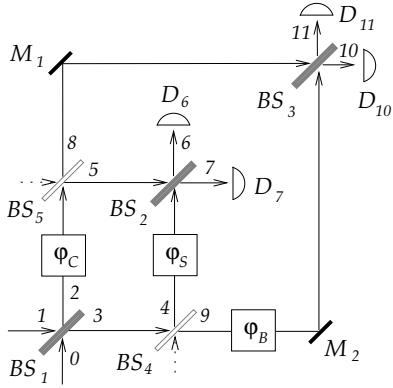}
\caption{Two braced Mach-Zehnder interferometers. $\varphi_C$
accounts for a voluntarily introduced delay common to both MZIs,
while the newly introduced $\varphi_S$ allows one to modify the
path length difference in the small MZI only.}
\label{fig:braced_MZIs_wave_nature_experiment}
\end{figure}

The beam splitter
$\text{BS}_2$ is added to the experimental setup as depicted in
Fig.~\ref{fig:braced_MZIs_wave_nature_experiment}.  We also add a
delay denoted $\varphi_S$, able to modify the path length
difference only in the ``small'' MZI, without affecting the
initial one. 
A detection at any of $D_6$ or $D_7$ cannot provide anymore the
which-path information for the second light quantum, therefore we
expect the interference to be restored.

We assume the state vector given by
Eq.~\eqref{eq:psi_input_state_11} at the input of $\text{BS}_1$.
In order to find the output state vector, we again express the
input field operators $\hat{a}_0^\dagger$ and $\hat{a}_1^\dagger$
in respect with the output ones. We obtain
\begin{eqnarray}
\label{eq:a0_dagger_in_respect_w_a6_a7_a10_a11_wave}
\hat{a}_0^\dagger=T_1\left(\left(T^2\text{e}^{i\varphi_C}+R^2\text{e}^{i\varphi_B}\right)\hat{a}_{10}^\dagger
+TR\left(\text{e}^{i\varphi_C}+\text{e}^{i\varphi_B}\right)\hat{a}_{11}^\dagger\right)
\nonumber\\
+R_1\left(TR\left(\text{e}^{i\varphi_C}+\text{e}^{i\varphi_S}\right)\hat{a}_{6}^\dagger
+\left(T^2\text{e}^{i\varphi_C}+R^2\text{e}^{i\varphi_S}\right)\hat{a}_{7}^\dagger\right)
\qquad
\end{eqnarray}
and
\begin{eqnarray}
\label{eq:a1_dagger_in_respect_w_a6_a7_a10_a11_wave}
\hat{a}_1^\dagger=T_1\left(TR\left(\text{e}^{i\varphi_C}+\text{e}^{i\varphi_B}\right)\hat{a}_{10}^\dagger
+\left(T^2\text{e}^{i\varphi_B}+R^2\text{e}^{i\varphi_C}\right)\hat{a}_{11}^\dagger\right)
\nonumber\\
+R_1\left(\left(T^2\text{e}^{i\varphi_S}+R^2\text{e}^{i\varphi_C}\right)\hat{a}_{6}^\dagger
+TR\left(\text{e}^{i\varphi_C}+\text{e}^{i\varphi_S}\right)\hat{a}_{7}^\dagger\right)
\qquad
\end{eqnarray}
If the beam splitters $\text{BS}_1$, $\text{BS}_2$ and
$\text{BS}_3$ are balanced,
after a series of long but straightforward calculations we obtain
the output state as
\begin{eqnarray}
\label{eq:psi_output_state_wave_nature} \vert\psi_{out}\rangle=
T_1^2\text{e}^{i(\varphi_C+\varphi_B)}\bigg(\frac{\sin\left(\Delta\varphi_B\right)}{\sqrt{2}}
\left(\vert2_{10}0_{11}\rangle-\vert0_{10}2_{11}\rangle\right)
 \nonumber\\
-\cos\left(\Delta\varphi_B\right)\vert1_{10}1_{11}\rangle\bigg)
+R_1^2\text{e}^{i(\varphi_C+\varphi_S)}\bigg(-\cos\left(\Delta\varphi_S\right)\vert1_{6}1_{7}\rangle
\nonumber\\  
+\frac{\sin\left(\Delta\varphi_S\right)}{\sqrt{2}}\left(
\vert0_62_{7}\rangle-\vert2_{6}0_7\rangle\right) \bigg)
\nonumber\\ 
+T_1R_1\text{e}^{i\frac{\varphi_B+\varphi_S}{2}}
 \left(
\cos\left(\frac{\Delta\varphi_B+\Delta\varphi_S}{2}\right)\vert1_{6}1_{10}\rangle
 \right.
 \nonumber\\
 \left.
+\sin\left(\frac{\Delta\varphi_B+\Delta\varphi_S}{2}\right)\vert1_{6}1_{11}\rangle
-\sin\left(\frac{\Delta\varphi_B+\Delta\varphi_S}{2}\right)\vert1_{7}1_{10}\rangle
\right.
\nonumber\\ 
\left.
+\cos\left(\frac{\Delta\varphi_B+\Delta\varphi_S}{2}\right)\vert1_{7}1_{11}\rangle
 \right)\qquad
\end{eqnarray}
where $\Delta\varphi_S=\varphi_S-\varphi_C$. For the probability
of coincident counts
$P_{10-11}=\vert\langle1_{10}1_{11}\vert\psi_{out}\rangle\vert^2$
we obtain again the result from
Eq.~\eqref{eq:P_coinc_10_11_particle}. The coincidence probability
at the inner MZI detectors yields now
\begin{eqnarray}
\label{eq:P_coinc_6_7_wave} P_{6-7}
=\vert\langle1_{6}1_{7}\vert\psi_{out}\rangle\vert^2
=\vert{R}_1\vert^4\cos^2\left(\Delta\varphi_S\right)
\end{eqnarray}
showing an interference pattern having the same spatial frequency
as the outer MZI, consistent with having two light quanta in the
interferometer.

This time however, the ``cross'' probabilities of coincident
counts also show interference patterns because no detection at any
of $D_6$ or $D_7$ can provide the which-way information. Indeed,
computing again the probability of coincident counts at detectors
$D_6$ and $D_{10}$ yields
\begin{eqnarray}
\label{eq:P_coinc_6_10_wave}
P_{6-10}
=\vert{T}_1R_1\vert^2\cos^2\left(\frac{\Delta\varphi_B+\Delta\varphi_S}{2}\right)
\end{eqnarray}
showing an interference pattern. However, the spatial frequency of
this interference pattern is halved compared to
Eqs.~\eqref{eq:P_coinc_10_11_particle} and
\eqref{eq:P_coinc_6_7_wave}, consistent with having one light
quantum in each MZI. This result is consistent with Ou \emph{et
al.} \cite{Ou90b} where they found the spatial frequency of the
interference fringes half the pump frequency.

It is noteworthy that in Eq.~\eqref{eq:P_coinc_6_10_wave} we have
$P_{6-10}\sim1/2\left(1+\cos\left(\varphi_B+\varphi_S-2\varphi_C\right)\right)$
\emph{i.e.} the delay element $\varphi_C$ ``sees'' two light
quanta.

\section{Delayed-choice quantum eraser}
\label{sec:delayed_choice_quantum_eraser} Similar to what was
performed in delayed choice experiments \cite{Jac07,Jac08,Hel87},
the beam splitters $\text{BS}_2$ and $\text{BS}_3$ and the
corresponding detectors can be put further apart, in order to
ensure space-like separation between detection events. Using
electro-optical couplers and polarized light for example
\cite{Jac07,Jac08}, we could ``insert'' and ``remove'' the beam
splitter $\text{BS}_2$ at will.

Focusing again on the experiment described in Section
\ref{sec:braced_MZIs_particle_nature}, detector $D_6$ can be
sufficiently far apart so that the detection  event would be
performed after the second light quantum left the MZI (that is
$\text{BS}_3$) and heading towards one of the detectors $D_{10}$
or $D_{11}$. Therefore, this (second) light quantum could not be
``informed'' about the ``collapse of the wavefunction'' due to a
detection event at $D_6$ and consequently its expected
particle-like behavior inside the MZI.

Moreover, the light quantum to be detected by $D_6$ might arrive
there after passing the beam splitter $\text{BS}_2$ (``inserted''
in a delayed-choice manner and being at a distance ensuring that
the insertion and detection events happen after the other light
quantum left $\text{BS}_3$). This is the experimental setup
discussed in Section \ref{sec:braced_MZIs_particle_wave}.
Therefore, the state inside the outer MZI should not ``collapse''
to $\vert0_81_9\rangle$ as discussed earlier, but would
``continue'' to be
$1/\sqrt{2}(\text{e}^{i\varphi_B}\vert0_81_9\rangle+\text{e}^{i\varphi_C}\vert1_80_9\rangle)$
implying interference fringes on varying the delays $\varphi_B$
and $\varphi_C$.

We could rightfully ask ourselves the question if any ``collapse''
is happening after all. To quote Englert, Scully and Walther
\cite{Eng99} ``a state reduction must be performed whenever we
wish to account for new information about the system''. In our
case, the detection event at $D_6$ (or $D_7$) yields this new
information, therefore the ``collapse of the wavefunction'' is
merely a mental process, not a physical one.

Finally, nobody ``informed'' the light quanta how to behave: the
experimenter(s) simply \emph{selected} and \emph{measured} a
certain point of view from our (quantum) system.

\section{Engineered states}
\label{sec:engineered_states} It is commonplace to think that if
we apply (classical or non-classical) light at the input of a MZI,
interference fringes will be present at its output on varying the
interferometer's arm length difference. 
But this has not to be always so, at least not for a
\emph{Gedankenexperiment}. We can create (non-classical) states of
light that yield no interference fringes. For example, if we apply
the input state
\begin{eqnarray}
\label{eq:psi_in_engineered_after_BS1_20_state}
\vert\psi_{in}\rangle=\frac{1}{2}\left(\vert2_00_1\rangle-\vert0_02_0\rangle-i\sqrt{2}\vert1_01_1\rangle\right)
\end{eqnarray}
to the beam splitter $\text{BS}_1$ we find
\begin{eqnarray}
\label{eq:psi_23_deterministic_engineered_state}
\vert\psi_{23}\rangle=\vert2_20_3\rangle
\end{eqnarray}
at its output, in other words we have \emph{with certainty} two
light quanta in the upper path and none in the lower one. This
state is fundamentally different from
Eq.~\eqref{eq:state_vector_after_BS1_balanced}, where we have the
two light quanta in a coherent superposition of being both in the
lower and in the upper arm of the interferometer. Using this input
state and
Eqs.~\eqref{eq:a0_dagger_in_respect_w_a6_a7_a10_a11_wave} and
\eqref{eq:a1_dagger_in_respect_w_a6_a7_a10_a11_wave} one gets the
output state
\begin{eqnarray}
\label{eq:psi_output_state_nointerference_eng}
\vert\psi_{out}\rangle=\frac{T_1^2}{2}\left(\vert2_{10}0_{11}\rangle-\vert0_{10}2_{11}\rangle+i\sqrt{2}\vert1_{10}1_{11}\rangle\right)
\nonumber\\
+\frac{R_1^2}{2}\left(-\vert2_{6}0_{7}\rangle+\vert0_{6}2_{7}\rangle+i\sqrt{2}\vert1_{6}1_{7}\rangle\right)
\nonumber\\
+T_1R_1\Big(i\vert1_{6}1_{10}\rangle+\vert1_{6}1_{11}\rangle+\vert1_{7}1_{10}\rangle+i\vert1_{7}1_{11}\rangle\Big)
\end{eqnarray}
showing indeed, no interference fringes although we are in a
``wave-like'' experimental setup and did not disturb in any way
the light quanta inside our interferometer. However, since
Eq.~\eqref{eq:psi_23_deterministic_engineered_state} allows one --
at least in principle -- to have the \emph{which-path knowledge},
interference cannot be present.

\begin{figure}
\centering
\includegraphics[width=3in]{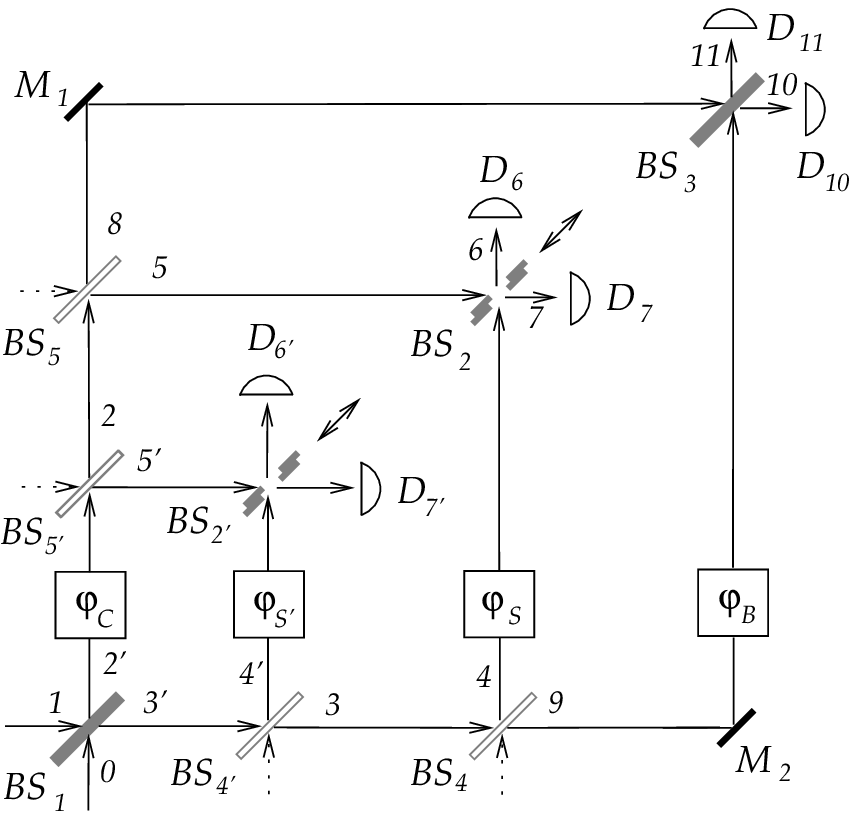}
\caption{Three braced Mach-Zehnder interferometers. The newly
introduced detectors $D_{6'}$ and $D_{7'}$ allow one to chose
between which-path information or wave-like behavior in the MZI
composed of $\text{BS}_1$, $\text{BS}_{2'}$, $\text{BS}_{4'}$ and
$\text{BS}_{5'}$. $\varphi_{S'}$ accounts for the path length
adjustments in this interferometer.}
\label{fig:triple_braced_MZIs_experiment}
\end{figure}

\section{Three or more braced interferometers}
\label{sec:Wigner_s_friend}
%
%
We consider now the extension of our previous experiments with a
new stage of beam splitters and detectors as depicted in
Fig.~\ref{fig:triple_braced_MZIs_experiment}. We further assume
that the newly introduced beam splitters $\text{BS}_{4'}$ and
$\text{BS}_{5'}$ are identical and are characterized by the
transmission (reflection) coefficients $T_{1'}$ ($ R_{1'}$). The
balanced (50/50) beam splitter $\text{BS}_{2'}$ can be inserted or
removed at will, allowing delayed choice experiments as discussed
earlier.

We need a ``high NOON'' state (with $N=3$ in this scenario) after
$\text{BS}_1$, therefore we create for the purpose of our
\emph{Gedankenexperiment} the input state,
\begin{eqnarray}
\label{eq:psi_in_engineered_after_BS1_30_03_state}
\vert\psi_{in}\rangle=\frac{1+i}{4}\Big(\vert3_00_1\rangle+\vert0_03_1\rangle
\qquad\qquad
\nonumber\\
\qquad\qquad
-\sqrt{3}\vert2_01_1\rangle-\sqrt{3}\vert1_02_1\rangle\Big)
\end{eqnarray}
so that after $\text{BS}_1$ the state vector is indeed
\begin{equation}
\label{eq:psi_N00N_state_BS1_30_03_state}
\vert\psi_{2'3'}\rangle=\frac{1}{\sqrt{2}}\left(\vert3_{2'}0_{3'}\rangle+\vert0_{2'}3_{3'}\rangle\right)
\end{equation}
We will focus only on the scenario when one of the detectors
$D_{6'}$ or $D_{7'}$ clicks once. If the experimenter measuring at
the detectors $D_{6'}-D_{7'}$ decides to erase the which-path
information (\emph{i.e.} $\text{BS}_{2'}$ is inserted), the state
after the beam splitters $\text{BS}_{4'}$ and $\text{BS}_{5'}$ is
again given by Eq.~\eqref{eq:state_vector_after_BS1_balanced} and
all scenarios described in Sections
\ref{sec:braced_MZIs_particle_nature} and
\ref{sec:braced_MZIs_particle_wave} remain available to the
experimenter(s) measuring at the detectors $D_{6}-D_{7}$ and
$D_{10}-D_{11}$. For example, if all experimenters agree to
measure wave-like properties, the triple coincidence probability
at detectors $D_{6'}$, $D_{6}$ and $D_{10}$ is found to be
\begin{eqnarray}
\label{eq:P_triple_coinc_P6_6prime_10}
P_{6'-6-10}=\frac{3}{4}\vert{T}_1T_{1'}^2R_1R_{1'}\vert^2
\qquad\qquad\qquad\qquad\qquad
\nonumber\\ 
\qquad
\cdot\Big(1+\sin\left(3\varphi_C-\varphi_B-\varphi_S-\varphi_{S'}\right)\Big)
\end{eqnarray}
showing indeed, interference fringes while varying any of the
delay elements. However, if the experimenter measuring at the
detectors $D_{6'}-D_{7'}$ decides to remove $\text{BS}_{2'}$ and
therefore obtain the which-path information, no interference could
be expected by correlating this measurement with the results
obtained by the other experimenter(s). Again, we can invoke the
``collapse of the wavefunction'' after $\text{BS}_{4'}$ and
$\text{BS}_{5'}$ to
Eq.~\eqref{eq:psi_23_deterministic_engineered_state} (assuming
that $D_{7'}$ clicked) or simply accept that we actually
\emph{selected} and \emph{measured} a particular point of view.

If the experimenter measuring at the detectors $D_{6'}-D_{7'}$ is
rather secretive and does not wish to reveal his/her experimental
setup, a density matrix approach as discussed in Appendix
\ref{app:density_matrix} applies to the double detection rates at
any of  $D_6$, $D_7$, $D_{10}$ and $D_{11}$. Therefore the
experimenter(s) measuring at detectors $D_6-D_7$ and
$D_{10}-D_{11}$ might grow frustrated that although the which-path
information is not available to them, their measurements show no
interference fringes whatsoever. Their error is that they ignore a
part of the maximally entangled wavevector given by
Eq.~\eqref{eq:psi_N00N_state_BS1_30_03_state}. Had they used as
input state Eq.~\eqref{eq:psi_input_state_11}, their ``cross''
probability of coincident counts (\emph{e.g.} $P_{6-10}$) would
have shown interference fringes\footnote{Of course, they could
also decide to register only triple coincidences (\emph{e.g.}
detections at $D_6-D_7-D_{10}$, $D_6-D_{10}-D_{11}$ etc.). The
secretive experimenter measuring at the detectors $D_{6'}-D_{7'}$
is therefore ``left out'' and his/her experimental setup plays no
role.}.

Extension to an even larger number of interferometers can be done
in a straightforward manner. If we use a carefully chosen state at
the input of $\text{BS}_1$ so that at its output the state is
\begin{eqnarray}
\label{eq:psi_high_NOON} \vert\psi_{{2^{(N-2)}}3^{(N-2)}}\rangle
=\frac{1}{\sqrt{2}}\Big(\vert{N}_{2^{(N-2)}}0_{3^{(N-2)}}\rangle
\qquad\qquad
\nonumber\\ 
 +\vert0_{2^{(N-2)}}N_{3^{(N-2)}}\rangle\Big)
\end{eqnarray}
with $N>3$ and we introduced in our experiment the beam splitters
$\text{BS}_{k'{}'}$, $\text{BS}_{k'{}'{}'}$ $\ldots$
$\text{BS}_{k^{(N-2)}}$ with $k=2,4,5$ and detectors $D_{j'{}'}$,
$D_{j'{}'{}'}$, $\ldots$ $D_{j^{(N-2)}}$ with $j=6,7$.

A $N^{th}$ order coincidence count at any detector pair
$D_{10}-D_{11}$, $D_6-D_7$, $D_{6'}-D_{7'}$, $\ldots$
$D_{6^{(N-2)}}-D_{7^{(N-2)}}$ can yield an interference pattern if
the corresponding beam splitter (\emph{i.e.} $\text{BS}_{3}$,
$\text{BS}_{2}$, $\text{BS}_{2'}$ etc.) is inserted. A cooperation
among $M\leq{N}$ experimenters can also yield interference for any
$N^{th}$ order coincident count\footnote{By ``$N^{th}$ order
coincident count'' we mean a coincident count among the $M$
cooperative users that adds up to $N$. For example in
Eq.~\eqref{eq:P_triple_coinc_P6_6prime_10} we had $N=3$ and three
experimenters, each one registering a single count.}, with the
constraint that all $M$ experimenters make a ``wave-like''
measurement. Moreover, a single secretive experimenter, not
wishing to share his/her information with the other experimenters
forces them to measure only statistical mixtures for any single,
double, triple, $\ldots$ $(N-1)^{th}$ order (coincidence) counts.

All these situations are summarized in Table~\ref{tab:1}.

\begin{table}
\caption{The various scenarios with $N$-photon states inside $N$
braced Mach-Zehnder interferometers (similar to
Fig.~\ref{fig:triple_braced_MZIs_experiment}); BCP stands for
``Bohr's complementary principle'' and ISK for ``incomplete system
knowledge''. Each experimenter controls a pair of detectors.}
\label{tab:1}       
\begin{tabular}{l|l|l|l}
 Path &  Cooperative & Non cooperative & Related  \\
 distingui- & experimenters & experimenters & con- \\
 \cline{2-3}
 shable? & \multicolumn{2}{|c|}{Interference pattern?} &  cepts \\
\noalign{\smallskip}\hline\noalign{\smallskip}
 yes, for & \multicolumn{2}{|c|}{no interference possible for any} & BCP \\
 all experi- & \multicolumn{2}{|c|}{single/coincident counts} & \\
 menters & \multicolumn{2}{|c|}{for any of the experimenters} & \\

\noalign{\smallskip}\hline\noalign{\smallskip}
 no, for  & for any $N^{th}$ order & only $N^{th}$ order & BCP/ \\
 none of & coincident counts & coincident counts & ISK \\
 the  expe- & distributed among  &  for a given& \\
 rimenters &  all experimenters & experimenter & \\

\noalign{\smallskip}\hline\noalign{\smallskip}
not for & for any $N^{th}$ order & only $N^{th}$ order & BCP/ \\
$M\leq N$  & coincident counts & coincident counts  & ISK \\
experi- &  distributed among & for a given experi-& \\
 menters  & $M$ experimenters & menter (among $M$)& \\


\end{tabular}
\end{table}
%

\section{Conclusions}
\label{sec:conclusions} In this paper we discussed various
experimental setups involving modified Mach-Zehnder
interferometers showing both the particle and the wave nature of
light. We were able to show that if a maximally entangled state is
used, the interference disappears if we \emph{select} events where
one light quantum gives the \emph{which-path information}. The
interference disappears \emph{without disturbing} in any way the
remaining light quantum (quanta). Interference can be revived if
we \emph{erase} the which-path information. A delayed-choice
version of the experiment can be performed, when the ``collapse of
the wavefunction'' inside one MZI would be determined by the
future decision on the erasure of a which-path information.
Extensions to three or more braced interferometers has also been
discussed.

As a final conclusion, Quantum Mechanics gives us a full and
coherent description of the results of our experiments, out of
which, the experimenter \emph{selects} a particular \emph{point of
view}. We did not need to invoke, at any point, the ``collapse of
the wavefunction'' picture to explain the obtained results.

\appendix

\section{Density matrix formalism approach}
\label{app:density_matrix} We assume two experimenters, the first
one controlling detectors $D_6-D_7$ and the second one detectors
$D_{10}-D_{11}$.

The following question might arise: if the first experimenter does
not wish to communicate his/her experimental setup, could the
second experimenter get this information while monitoring the
detection rates at $D_{10}$ and/or $D_{11}$?

In order to answer this question, we shall use the density matrix
approach. Assuming the experimental setup from
Fig.~\ref{fig:Braced_MZIs_particle_nature_experiment}, the output
state vector is given by
Eq.~\eqref{eq:psi_output_state_particle_nature} therefore we have
the output density matrix
\begin{equation}
\hat{\rho}_{out}=\vert\psi_{out}\rangle\langle\psi_{out}\vert
\end{equation}
If we are forced to ignore detectors $D_6$ and $D_7$, the output
density matrix $\hat{\rho}_{out}$ has to be traced over the modes
$6$ and $7$ yielding the reduced density matrix
\begin{equation}
\label{eq:rho_reduced_density_analytic}
\hat{\rho}_{10,11}=\text{Tr}_{6,7}\left\{\hat{\rho}_{out}\right\}
=\sum_{m,n=0}^{\infty}{\langle{m}_6n_7\vert\psi_{out}\rangle\langle\psi_{out}\vert{m}_6n_7\rangle}
\end{equation}
and after some calculations we arrive at the expression
\begin{eqnarray}
\label{eq:rho_reduced_6_7_2MZI}
\hat{\rho}_{10,11}=\frac{\vert{T_1}\vert^4}{2}\sin^2\left(\Delta\varphi_B\right)
\Big(\vert2_{10}0_{11}\rangle\langle2_{10}0_{11}\vert
 \qquad\qquad
\nonumber\\ 
+\vert0_{10}2_{11}\rangle\langle0_{10}2_{11}\vert\Big)
+\vert{T_1}\vert^4\cos^2\left(\Delta\varphi_B\right)\vert1_{10}1_{11}\rangle\langle1_{10}1_{11}\vert
\nonumber\\ 
+\vert{R_1}\vert^4\vert0_{10}0_{11}\rangle\langle0_{10}0_{11}\vert
+\vert{T_1}R_1\vert^2\Big(\vert1_{10}0_{11}\rangle\langle1_{10}0_{11}\vert
\nonumber\\ 
+\vert0_{10}1_{11}\rangle\langle0_{10}1_{11}\vert\Big) \qquad
\end{eqnarray}
We are now able to compute any photo-count probability at $D_{10}$
and/or $D_{11}$. For example, computing the probability of
coincident counts at the detectors $D_{10}$ and $D_{11}$ we get
\begin{eqnarray}
\label{eq:P_coinc_10_11_density_matrix}
P_{10-11}=\text{Tr}\left\{\hat{a}_{10}^\dagger\hat{a}_{10}\hat{a}_{11}^\dagger\hat{a}_{11}\hat{\rho}_{10,11}\right\}
=\vert{T}_1\vert^4\cos^2\left(\Delta\varphi_B\right)\quad
\end{eqnarray}
and we revisit the result from
Eq.~\eqref{eq:P_coinc_10_11_particle}. However, one could
speculate that the single detection rates at any of the detectors
$D_{10}$ or $D_{11}$ could yield some supplementary information.
Therefore, we compute the single detection rate at, for example,
detector $D_{10}$, yielding
\begin{eqnarray}
\label{eq:P_single_10_single_density_matrix_appr}
P_{10}=\text{Tr}\left\{\hat{a}_{10}^\dagger\hat{a}_{10}\hat{\rho}_{10,11}\right\}
=\vert{T}_1\vert^4+\vert{T}_1R_1\vert^2
\end{eqnarray}
and no interference fringe variation can be found by varying any
arm length difference. But we could blame this result on the
``particle-like'' experimental setup from
Fig.~\ref{fig:Braced_MZIs_particle_nature_experiment}.

Therefore, we now assume that the experimenter measuring at
detectors $D_{6}-D_{7}$ changed his/her setup according to
Fig.~\ref{fig:braced_MZIs_wave_nature_experiment}. We have to
compute the density matrix starting from
Eq.~\eqref{eq:psi_output_state_wave_nature} and again, trace over
the inner detectors $D_6$ and $D_7$. After computing the partial
trace from Eq.~\eqref{eq:rho_reduced_density_analytic} we get
\emph{the same result} Eq.~\eqref{eq:rho_reduced_6_7_2MZI} for
$\hat{\rho}_{10,11}$. Therefore, in the ``wave-like'' experimental
setup we find the same single detection rate $P_{10}$ given by
Eq.~\eqref{eq:P_single_10_single_density_matrix_appr}. We can
discard now the speculation done previously: no information about
the experimental setup at the inner detectors $D_6$ and $D_7$ can
be retrieved from any measurement at the outer MZI.



One could wonder if this result is consistent. The experimenter at
the inner detectors has no which-path information, therefore the
dictum ``ignorance is interference'' should apply. However, from
Eq.~\eqref{eq:P_single_10_single_density_matrix_appr} it is clear
that it does not.

Actually, by ignoring the fact that we started from an
\emph{entangled state} and considering only partial measurements,
we are bound to find only statistical mixtures. The state vector
$\vert\psi_{out}\rangle$ describes the whole system (\emph{i.e.}
both experimenters) and only the knowledge of their \emph{global
state} can yield the interference fringes from
Eq.~\eqref{eq:P_coinc_6_10_wave}.

%
%

\end{document}